# Patient-perceived progression in multiple system atrophy: natural history of quality of life

Tiphaine Saulnier,[1] Margherita Fabbri,[2,3] Mélanie Le Goff,[1] Catherine Helmer,[1,4] Anne Pavy-Le Traon,[2,3] Wassilios G. Meissner,[6,7,8] Olivier Rascol,[2,3] Cécile Proust-Lima,[1,4,†] and Alexandra Foubert-Samier[1,4,6,7,†]

[†]**These authors contributed equally to this work.**

# Abstract

Health-related quality of life (Hr-QoL) scales provide crucial information on neurodegenerative disease progression, help improving patient care, and constitute a meaningful endpoint for therapeutic research. However, Hr-QoL progression is usually poorly documented, as for multiple system atrophy (MSA), a rare and rapidly progressing alpha-synucleinopathy. This work aimed to describe Hr-QoL progression during the natural course of MSA, explore disparities between patients, and identify informative items using a four-step statistical strategy.

We leveraged the data of the French MSA cohort comprising annual assessments with the MSA-QoL questionnaire for more than 500 patients over up to 11 years. The four-step strategy (1) determined the subdimensions of Hr-QoL in MSA; (2) modelled the subdimension trajectories over time, accounting for the risk of death; (3) mapped the sequence of item impairments with disease stages; and (4) identified the most informative items specific to each disease stage.

Among the 536 patients included, 50% were women and they were aged on average 65.1 years old at entry. Among them, 63.1% died during the follow-up. Four dimensions were identified. In addition to the original motor, nonmotor, and emotional domains, an oropharyngeal component was highlighted. While the motor and oropharyngeal domains deteriorated rapidly, the nonmotor and emotional aspects were already slightly to moderately impaired at cohort entry and deteriorated slowly over the course of the disease. Impairments were associated with sex, diagnosis subtype, and delay since symptom onset. Except for the emotional domain, each dimension was driven by key identified items.

Hr-QoL is a multidimensional concept that deteriorates progressively over the course of MSA and brings essential knowledge for improving patient care. As exemplified with MSA, the thorough description of Hr-QoL using the 4-step original analysis can provide new perspectives on neurodegenerative diseases' management to ultimately deliver better support focused on the patient's perspective.

**Author affiliations:**

1 Univ. Bordeaux, Bordeaux Population Health Research Center, Inserm U1219, F-33000 Bordeaux

2 MSA French Reference Center, Univ. Hospital Toulouse, F-31000 Toulouse

3 Univ. Toulouse, CIC-1436, Departments of Clinical Pharmacology and Neurosciences, NeuroToul COEN Center, NS-Park/FCRIN Network, Toulouse University Hospital, Inserm U1048/1214, F-31000 Toulouse

4 Inserm, CIC1401-EC, F-33000 Bordeaux

5 ToNIC, Toulouse NeuroImaging Center, Univ Toulouse, Inserm, UPS, F-31000 Toulouse, France

6 CHU Bordeaux, Service de Neurologie des Maladies Neurodégénératives, IMNc, CRMR AMS, NS-Park/FCRIN Network, F-33000 Bordeaux, France

7 Univ. Bordeaux, CNRS, IMN, UMR5293, F-33000 Bordeaux

8 Dept. Medicine, University of Otago, Christchurch, and New Zealand Brain Research Institute, Christchurch, New Zealand

Correspondence to: Cécile Proust-Lima

146 rue Léo Saignat, F-33000 Bordeaux, France

cecile.proust-lima@inserm.fr

**Running title**: Quality of life progression in MSA

# Introduction

Multiple system atrophy (MSA) is a rare, neurodegenerative, and incurable disease characterized by a variable combination of parkinsonism, cerebellar impairment, and autonomic disorders. The disease has a progressive and rapid global degradation, and a poor prognosis. In MSA as in other neurodegenerative diseases, health-related quality of life (Hr-QoL) is rapidly affected[1–3] and strongly related to the disease process. Therefore, studying the disease progression with a focus on patients' perception can provide crucial information on the disease course, giving clinicians opportunities to deliver better support.[4–6] In recent years, Hr-QoL has been a key domain in the study of Parkinson's disease used to adapt care planning.[6–8] Supported by the World Health Organization's (WHO) and the Food and Drug Administration (FDA), the assessment of Hr-QoL has become crucial for the improvement of care and drug development.[9] However, the progression of Hr-QoL in MSA remains insufficiently documented.[4,6,8,10] Beyond the disease rarity and availability of large cohorts, this deficit is explained by the statistical challenges raised by Hr-QoL data.

Hr-QoL is a complex concept, reflecting multiple aspects such as physical condition, psychological state, and social relationships. It is usually assessed by Likert measurement scales composed of numerous items transcribing the disease manifestations experienced by patients with graded scores.[6,11] Each item provides relevant information that needs to be accounted for, which prevents the use of sum scores.[12] Moreover, the study of changes in Hr-QoL over time requires the use of statistical methods adapted to longitudinal data and to occurrence of events such as death, which interrupt the follow-up inducing missing data, usually for patients with highest impairments.[13] Finally, in neurodegenerative diseases, the progression of Hr-QoL would benefit from being mapped against clinical progression to better understand the link between clinical and QoL impairments.

Based on the French MSA cohort, this work aimed to better understand QoL evolution during the natural disease course, identify factors associated with progression, and relate progression to disease staging. For this purpose, we addressed the methodological challenges with a four-step statistical strategy to analyse longitudinal Hr-QoL data, applicable to any neurodegenerative disease.

# Materials and methods

## Study population and materials

The French MSA cohort was created in 2007 by the French Reference Centre, a collaboration between the University Hospitals in Bordeaux and Toulouse.[1] Its constitution has been registered with the CNIL (Commission Nationale Informatique et Liberté). It is an open and prospective cohort that includes all consenting patients diagnosed with MSA according to the Gilman criteria[14] and consists of annual follow-ups with standardized clinical assessment. For this work, all inclusion and follow-up data prior to December 31, 2021 (called administrative censoring), were considered.

### Ascertainment of MSA-QoL progression

Hr-QoL was assessed by the MSA-QoL questionnaire, developed in 2008.[6,11] The MSA-QoL is composed of 40 ordinal items (Table 2), each with five increasing levels of impairment (0 no, 1 slight, 2 moderate, 3 marked, and 4 extreme impairment). The scale assesses three QoL domains: motor, nonmotor, and emotional/social. The original version was translated to adapt to the French-speaking audience.[15] Since 2008, patients have been asked to complete the MSA-QoL questionnaire during each annual consultation.

### Ascertainment of MSA progression

MSA progression was assessed at each visit by the Unified MSA Rating Scale (UMSARS)[16] part IV, a global disability score that distinguishes among five stages (completely independent (stage I), needs help with some chores (stage II), needs help with half of the chores (stage III), does a few chores alone (stage IV), and totally dependent (stage V)).[5,6]

## Ascertainment of death and dropout

At the time of administrative censoring, patients were either classified as deceased, still alive in the cohort, or dropped out if their last visit was prior to July 1st, 2020 (i.e., more than 1.5 years before administrative censoring).

## Ascertainment of associated factors

We considered the following factors that may influence Hr-QoL or personal perceptions: sex, age at inclusion, delay between symptom onset and inclusion, predominant syndrome (parkinsonian or cerebellar)[14], diagnosis certainty (possible or probable), and presence of orthostatic hypotension (decrease greater than 10 mmHg in diastolic or greater than 20 mmHg in systolic blood pressure between the supine and upright positions) or urinary disorder (UMSARS-I Item 10 score > 2) at inclusion.[5,6] Treatment effects on Hr-QoL were explored for L-dopa (motor dimension), antihypotensive agents (nonmotor), and antidepressants (emotional/social).

## Sample selection

We included all patients who had completed at least one MSA-QoL questionnaire before administrative censoring and without missing data for all factors listed previously (Supplementary Fig. S1).

# Statistical analyses

The longitudinal statistical analysis of Hr-QoL was divided into four steps to successively (1) identify the homogeneous dimensions within the scale, (2) model each dimension's trajectory and explore disparities between patient profiles, (3) map items' impairment hierarchy with the course of the disease, and (4) identify the most informative items at each disease stage. Steps are briefly described below (see Supplementary Appendix 1 for details).

## Step 1: Identification of homogeneous subdimensions of MSA-QoL

The different dimensions measured by the scale were identified using factorial analyses, following the PROMIS methodology.[17] The objective was to ensure that all items from a subdimension measured the same phenomenon (*unidimensionality*), did not carry redundant information (*conditional independence*), and higher levels of items always corresponded to higher levels of QoL impairment (*increasing monotonicity*). This step, necessary to ensure the validity of the statistical analyses in the subsequent steps, was carried out on all observed repeated individual follow-up data, representing 1537 MSA-QoL questionnaires for 557 patients.

Steps 2 to 4 were performed separately on each dimension. All patients with at least one item completed per (modified-)dimension were included, leading to a sample of 536 patients with 1501 visits for Step 2, and with at least 75% of the MSA-QoL items completed per (modified-)dimension, leading to a sample of 516 patients with 1376 visits for Step 3.

## Step 2: Description of MSA-QoL item trajectories over time and their associated factors

The trajectory of each dimension continuum was modelled over time using a longitudinal item response theory (longIRT) model for repeated graded item responses.[18,19] This model combined a linear mixed model to describe the underlying dimension deterioration over time according to covariates with cumulative probit measurement models to define the link between the underlying dimension and each item observation. To account for the informative dropout induced by early deaths, the instantaneous risk of death was simultaneously modelled according to the current dimension level within a joint model.[13] Other dropouts were assumed missing at random. The linear mixed model included a linear function of time since inclusion at the population and individual levels and was adjusted for covariates as simple effects to explore phenotype differences according to sex, predominant syndrome, diagnosis certainty, age at inclusion, presence of orthostatic hypotension or urinary disorder at inclusion, delay since symptom onset, and treatments. Time-dependent binary treatments were considered for the associated dimensions.

## Step 3: Mapping of item impairment hierarchy to disease stages

The sequence of item impairments derived from Step 2 was defined according to the dimension-specific continuum and could not be overlaid across scale subdimensions. Step 3 consisted of anchoring each dimension continuum to the disease stage to improve the understanding of the sequences. This was achieved by jointly modelling the repeated data of a subdimension sum score with the repeated data of the disease stage (in a longIRT model) and determining the level of each dimension continuum that corresponded to a change in disease stage.

## Step 4: Listing of the most informative items by disease stage

None of the items contributed uniformly within or across disease stages. The contribution of each item was quantified through the percentage of information it carried at each stage (i.e., the Fisher information the item carried standardized by the total Fisher information for a given disease stage).[20,21] The most informative items were identified as those carrying the highest percentages of information at several disease stages.

# Results

## Demographics

Among the 536 patients, 50% were women, 57.5% were from Bordeaux, 67.7% were diagnosed with MSA-P, and 74.6% with probable certainty (Table 1). Patients were on average 60.6 years old at symptom onset and 65.1 years old at inclusion, with a delay since symptom onset of approximately 4.5 years. At inclusion, 67.4% of patients had orthostatic hypotension, and 68.1% had a urinary disorder. At inclusion, 68.3% were taking L-dopa, 30.0% antihypotensive agents, and 20.7% antidepressants. A total of 1501 follow-up visits were analysed, representing approximately 2.8 observations per patient. During follow-up, we recorded 63.1% deaths and 16.2% dropouts.

## Identification of four MSA-QoL subdimensions

The PROMIS methodology confirmed the three dimensions previously identified by Schrag et al[11] (Table 2) with motor, nonmotor, and emotional aspects, but it also identified a fourth dimension featuring oropharyngeal impairment as assessed by items 6, 7, 9, and 10. The distribution of the items within each dimension was identical to the original distribution, except for Item 15 (bladder problems) which moved from the original nonmotor dimension to the modified motor dimension. Items 25 (slowness of thinking) and 26 (difficulties concentrating) were strongly correlated providing redundant information. We thus removed Item 25 from the modified scale. In the end, the modified MSA-QoL scale assessed four QoL dimensions: motor (10 items), oropharyngeal (4 items), nonmotor (11 items), and emotional/social (14 items).

## Item trajectories and sequences and associated factors

Most of the items already showed partial impairment at inclusion for the reference patient profile (Supplementary Fig. S2), with the exception of the oropharyngeal items (swallowing, feeding and drinking). The items in the motor dimension deteriorated very rapidly over the first 4 years of follow-up, with most of the items reaching level 3 (marked impairment) less than 5 years after inclusion. Moving, walking, balance, housework and hobbies were already perceived as moderately impaired (level 2) at inclusion. Dressing and requiring help to go to the toilet were moderately impaired by 3-4 years after inclusion. The oropharyngeal items progressively deteriorated, reaching moderate impairment approximately 5 years after inclusion and marked impairment by 10 years in the reference patient category. Drooling was reported to rapidly increase. In comparison, the nonmotor and emotional spheres deteriorated very slowly. The most impacted items (with a roughly moderate impairment at entry) were lack of energy, fatigue, and constipation in the nonmotor dimension and participation in social activities and concerns about the future in the emotional/social dimension.

The sequence of the item impairments according to each subdimension continuum was derived from the model (Figure 1). Within the motor dimension, balance (Item 3), handwriting (8), and bladder (15) problems occurred first, followed by difficulties walking (2), speaking (5), and continuing hobbies (13). Difficulties going to the toilet arose later than other motor impairments but progressed rapidly. In the oropharyngeal dimension, drooling occurred first (7). In the nonmotor dimension, fatigue (23), lack of energy (24), and constipation (16) occurred first, followed by localized pain (18, 19, and 20) and night discomfort (21). In the emotional/social dimension, patients first felt incapable (30), worried about their future (31) and their family (32), gave up social activities (37) and were embarrassed to talk to people (39). Then, they gradually lost their motivation (29), their confidence (34) and felt bored (40), frustrated (27), and depressed (28) to the point of breaking ties with their friends (36) and isolating (33).

Each subdimension continuum was modulated by other factors (Figure 1). The motor and oropharyngeal dimensions deteriorated much more rapidly than the nonmotor and emotional dimensions. Female patients had poorer Hr-QoL than males in the four dimensions. Patients diagnosed with MSA-P had significantly lower motor but higher oropharyngeal, nonmotor and emotional/social deteriorations on average than patients diagnosed with MSA-C. In the nonmotor dimension, the difference in impairment between MSA-P and MSA-C patients was equivalent to the effect of 2 years of progression. There were no noticeable differences in diagnosis certainty, age, or the presence of hypotension at inclusion. Patients suffering from urinary disorders at inclusion had a lower Hr-QoL according to the motor, oropharyngeal, and nonmotor dimensions. The delay since symptom onset was associated with higher Hr-QoL impairment in the motor, oropharyngeal, and emotional/social dimensions. Finally, Hr-QoL impairment appeared higher in the motor dimension for patients taking L-dopa and in the nonmotor dimension for patients under antihypotensive treatment.

## Mapping quality of life deterioration to disease stages (UMSARS-IV)

Projections of MSA disease stages onto the sequence of MSA-QoL impairments (Figure 1) showed that motor and oropharyngeal dimensions were not affected during Stage I but deteriorated very quickly over the subsequent stages with all their items reaching the maximum level by the beginning of Stage V. The degradation of nonmotor and emotional/social dimensions was more progressive, occurring over the course of the disease.

## Most informative items over the disease course

The percentage of information carried by each item within the five MSA disease stages (Figure 2 and Supplementary Table S1) allowed the identification of the most informative items during the course of the disease.

In the motor dimension, since all items had already reached the maximum level at the beginning of Stage V (Figure 1), the most contributing items were identified from the 4 first stages only. They were as follows: 1 (moving), 2 (walking), 4 (standing up), 11 (dressing), 12 (toilet), and 14 (housework). Item 12, toilet, was poorly informative during Stage I but captured an increasing proportion of information during Stages II-IV (9%, 14%, and 17%, respectively), as well as Item 4, standing up, and Item 11, dressing.

In the oropharyngeal dimension, Item 7, saliva, carried 58% of the information at Stage I but became secondary at later stages. At Stages II to V, Items 6 (swallowing), 9 (feeding), and 10 (drinking) became major providing together more than 85% of stage-specific information.

In the nonmotor dimension, Items 23 (tired) and 24 (energy) together carried more than 51%, 36%, 35%, and 33% of the information for Stages I, II, III, and IV, respectively. Items 19 (neck/shoulder pain), 20 (leg/back pain), and 21 (comfortable) carried a large part of the information for Stages II to V (more than 36%). In contrast, Items 17 (dizziness), 18 (cold hands/feet), 22 (breathing), and 26 (concentration) were less relevant in the early disease stages.

Finally, in the emotional/social dimension, the information was more equally spread across items: no item seemed to concentrate all emotional Hr-QoL information regardless of stage. The four main items, 27 (being frustrated), 28 (depressed), 29 (loss of motivation), and 40 (feeling bored), carried slightly more information than the others over all stages (with more than 8% of information each at Stages II, III and IV).

# Discussion

By leveraging Hr-QoL data from the French MSA cohort, we proposed a patient perspective on disease progression to improve MSA management. First, the evaluation of the MSA-QoL confirmed the original scale structure and identified a fourth dimension based on four oropharyngeal items. This additional subdimension is particularly significant, as feeding aspects are strongly impacted by MSA progression, with swallowing disorders being a major risk for death. Additionally, the reclassification of Item 15 suggests that bladder dysfunction, rather than motor impairment, greatly affects patients' ability to move to the toilet. Second, the dimensions exhibited distinct patterns of deterioration throughout the disease course. The motor and oropharyngeal dimensions showed minimal impairment during Stage I but deteriorated rapidly thereafter. In contrast, the nonmotor and emotional/social dimensions were slightly to moderately impaired at Stage I and progressed slowly. Third, Hr-QoL impairments varied across patient profiles. Female patients experienced poorer overall Hr-QoL than males. MSA-P patients showed lower motor deterioration but higher oropharyngeal, nonmotor, and emotional/social deterioration than MSA-C patients. These findings align with Xiao *et al.*[10] study. The presence of urinary disorders at inclusion and the delay since symptom onset influenced impairment, emphasizing the need for early diagnosis. Patients receiving L-dopa and antihypotensive treatment exhibited higher Hr-QoL impairments in the motor and nonmotor dimensions respectively, suggesting either their limited effectiveness on QoL-related symptoms or their systematic prescription to the most affected patients.

This study identified the most informative Hr-QoL items, providing guidance to clinicians regarding the most critical domains. This information could facilitate personalized clinical attention and management. Throughout disease progression, key features emerge. In Stage I, a more precise evaluation and treatment of drooling, particularly for MSA-P patients, appeared beneficial. Additionally, early evaluation and treatment of urinary dysfunction was associated with better Hr-QoL. In Stages II and III, the impact on activities of daily living and self-care, combined with increased loss of mobility, substantially contributes to QoL. Prioritizing the implementation of technical and human assistance along with sustained rehabilitation to maintain autonomy appears crucial. Fatigue remained a key element throughout all stages of the disease. As reported in other studies[22], potential determinants such as orthostatic hypotension, sleep disorders, or depressive symptoms can contribute to the fatigue reported by patients. Although orthostatic hypotension was not a major predictor, antihypotensive treatment was significantly associated with a higher impact of nonmotor symptoms, such as fatigue and loss of energy. While the impact of sleep disorders was not evaluated, they are a common occurrence in MSA[23] and may contribute to fatigue. Additionally, throughout the entire disease course, psychological support, with or without antidepressant treatment, appears crucial with early attention to self-esteem and future outlook, followed by consideration of the disease's impact on social and family interactions. All recommendations to enhance the management of Hr-QoL in MSA are summarized in a mind map (Figure 3).

The strength of this study lies in the use of one of the largest MSA cohort worldwide, with extensive duration of follow-up and high-quality data combined with an original statistical strategy. The four-step statistical strategy addressed the challenges posed by repeated Hr-QoL data by decomposing the scale into unidimensions, accounting for the informative higher risk of death during the follow-up when describing each dimension's trajectory, mapping the impairment hierarchy of Hr-QoL items with disease stages, and identifying the most informative items.

In conclusion, describing the natural history of Hr-QoL in MSA through an innovative statistical approach provided practical recommendations for the management of MSA patients. The same methodology could be replicated in other neurodegenerative diseases to improve disease understanding and management from the patient's perspective.

## Data availability

Anonymized data can only be shared by reasonable motivated request to the MSA reference centre coordination.

## Acknowledgements

We would like to thanks the French National Research Agency (Project DyMES - ANR-18-C36-0004-01) and the Nouvelle-Aquitaine region (Project AAPR2021A-2020-11937310) for their financial support that made this work possible. Several authors of this publication are members of the European Reference Network for Rare Neurological Diseases (Project ID No 739510).

## Funding

This work was funded by the French National Research Agency (DyMES-ANR-18-C36-0004-01), and the region Nouvelle-Aquitaine (AAPR2021A-2020-11937310).

## Competing interests

TS, ML, CH, and CPL have nothing to disclose.

MF received honoraria to speak from BIAL, AbbVie, Orkyn, Consultancies for Bial and LVL Médical.

APL received honoraria from Almylam and Biohaven.

PP received honoraria from IKT Laboratory and grant from the French Research Agency.

WGM has received fees for editorial activities with Elsevier, and consultancy fees from Lundbeck, Biohaven, Roche, Alterity, Servier, Inhibikase, Takeda, and Teva.

OR received honoraria for scientific advises from companies developing novel therapies of biomarkers for MSA, including Lundbeck, Neuralight, ONO Pharma, Servier, Takeda, UCB and received scientific grants for MSA from the French

# Supplementary material

Supplementary material is available at *Brain* online.

# Figure legends

**Figure 1: Sequence of Hr-QoL item impairment according to disease stage and associated factors for each subdimension.** Each graph represents the items' degradation according to the underlying subdimension continuum. The color gradient reflects increasing impairment, i.e., the lightest color for level 0 ("No problem") to the darkest color for level 4 ("Extreme problem"). The intensity of the change in the underlying continuum due to each factor is represented by the horizontal black bar. The significance of a covariate effect is denoted by '°' for $p \leq 0{\cdot}1$, '*' for $p \leq 0{\cdot}05$, '**' for $p \leq 0{\cdot}01$, and '***' for $p \leq 0{\cdot}001$ (very significant) with $p$ the associated Wald-test p-value. The four vertical black lines indicate the estimated location of the five MSA disease stages on each subdimension continuum. The different reference categories are male, MSA-C subtype, diagnosis with possible certainty, no orthostatic hypotension or urinary disorder at entry, no delay since symptom onset, and no treatment (among L-dopa, antihypotensive agents, and antidepressants).

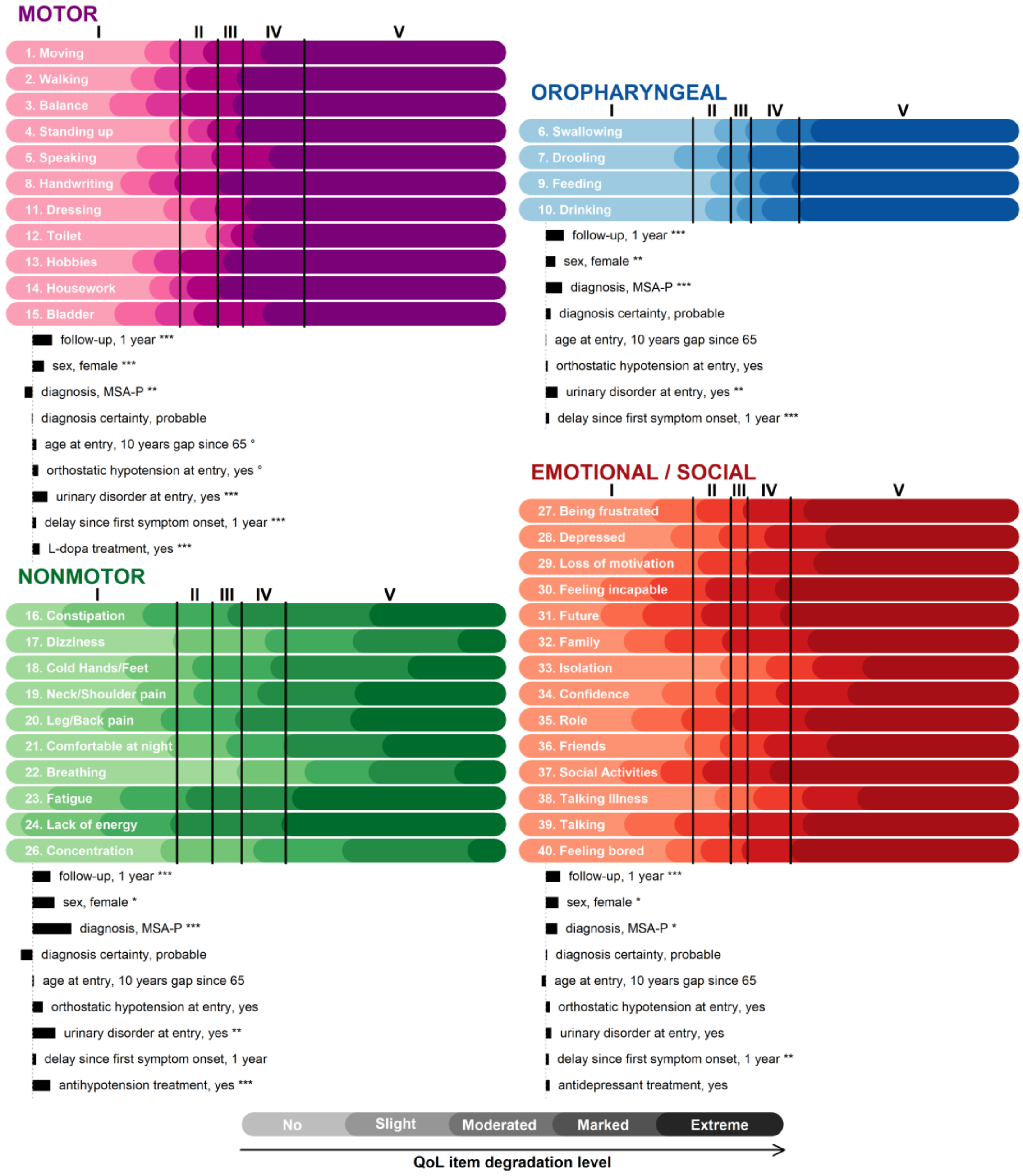

**Figure 2: Item contribution for each subdimension across the five MSA disease stages.** The larger the color bar, the more the item contributed. Grey bar plots report the total information carried by the items for each subdimension to the five disease stages.

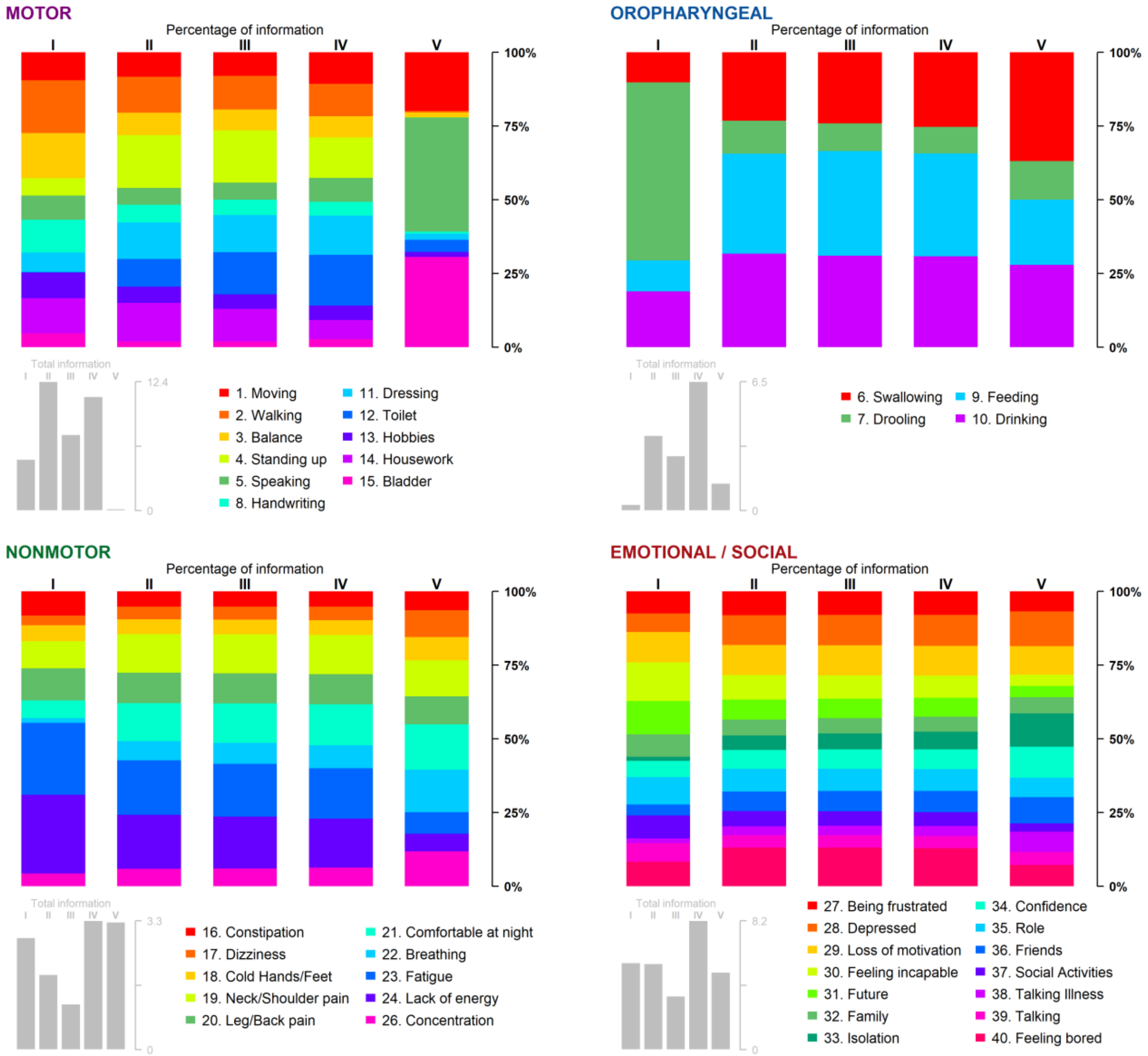

**Figure 3: Pathway for improving Hr-QoL management in MSA.**

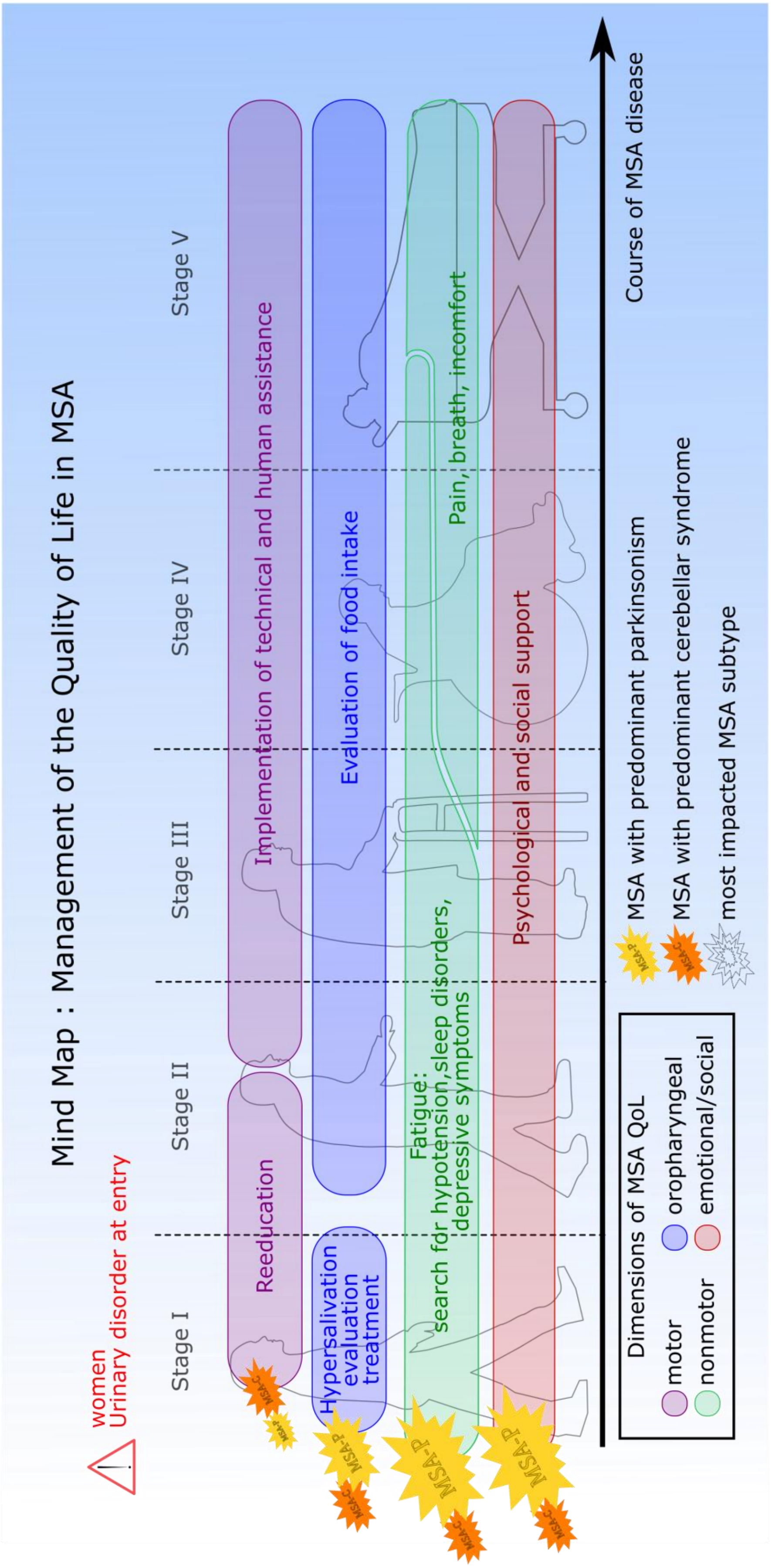